%% file: main.tex
\documentclass[dvipsnames,screen,sigplan,pbalance]{acmart}

\usepackage{xcolor}
\usepackage{nameref}
\usepackage{hyperref}
\usepackage[nameinlink]{cleveref}
\usepackage{csvsimple, siunitx, ifthen}

\usepackage{semantic}   
\usepackage{braket}     
\usepackage{natbib}
\usepackage{subcaption}
\usepackage{tikz}
\usetikzlibrary{arrows.meta}
\usetikzlibrary{patterns}
\usepackage{pgfplots}
\usepackage{pgfplotstable}
\usepackage{mathpartir}
\usepackage{enumitem}
\usepackage{stmaryrd}
\usepackage{marginnote}

\usepackage{cancel}
\usepackage{setspace}
\usepackage{pdfpages}
\usepackage{multirow, diagbox, makecell}
\usepackage{doi}
\usepackage{minted} \BeforeBeginEnvironment{minted}{\begingroup\color{black}} \AfterEndEnvironment{minted}{\endgroup} \setminted{autogobble,breaklines,breakanywhere,linenos, numbersep=5pt, xleftmargin=7pt}
\usepackage{listings}
\usepackage{xspace}
\usepackage{esvect}
\usepackage{letltxmacro}
\usepackage{microtype}
\LetLtxMacro{\rulelabel}{\label}
\LetLtxMacro{\lemmalabel}{\label}

\tikzset{
  hatch distance/.store in=\hatchdistance,
  hatch distance=10pt,
  hatch thickness/.store in=\hatchthickness,
  hatch thickness=2pt
}

\Crefname{Requirement}{Requirement}{Requirement}

\newcounter{reqcounter}

\makeatletter

\newcommand{\judgmath}[1]{\(#1\)}
\newcommand{\judgshape}[2][]{\begin{flushleft}\fbox{\judgmath{#2}~#1}\end{flushleft}}

\newcommand{\techprefix}{}
\newcommand{\rulename}[1]{{\scshape #1}}

\newcommand{\@defruleStar}[3][\techprefix]{\phantomsection{\rulename{#3}}\expandafter\rulelabel{rule:#1:#2}}
\newcommand{\@defruleNoStar}[2][\techprefix]{\@defruleStar[#1]{#2}{#2}}
\newcommand{\defrule}{\@ifstar\@defruleStar\@defruleNoStar}

\newcommand{\@refruleStar}[3][\techprefix]{\hyperref[rule:#1:#2]{Rule \rulename{#3}}}
\newcommand{\@refruleNoStar}[2][\techprefix]{\@refruleStar[#1]{#2}{#2}}
\newcommand{\refrule}{\@ifstar\@refruleStar\@refruleNoStar}

\makeatother

\newcommand\hcancel[2][black]{\setbox0=\hbox{$#2$}%
\rlap{\raisebox{.45\ht0}{\textcolor{#1}{\rule{\wd0}{1pt}}}}#2}

\newcommand{\materials}{software artifact \cite{artifact}\xspace}

\setcopyright{cc}
\setcctype{by}
\acmDOI{10.1145/3759537.3762696}
\acmYear{2025}
\copyrightyear{2025}
\acmISBN{979-8-4007-2162-5/25/10}
\acmConference[Scheme '25]{Proceedings of the 26th ACM SIGPLAN International Workshop on Scheme and Functional Programming}{October 12--18, 2025}{Singapore, Singapore}
\acmBooktitle{Proceedings of the 26th ACM SIGPLAN International Workshop on Scheme and Functional Programming (Scheme '25), October 12--18, 2025, Singapore, Singapore}
\received{2025-07-24}
\received[accepted]{2025-08-14}

\begin{document}
\title{Fast and Extensible Hybrid Embeddings with Micros}

\author{Sean Bocirnea}
\orcid{0009-0007-5231-8618}
\affiliation{%
  \institution{University of British Columbia}
  \city{Vancouver}
  \country{Canada}
}
\email{seanboc@cs.ubc.ca}

\author{William J. Bowman}
\orcid{0000-0002-6402-4840}
\affiliation{%
  \institution{University of British Columbia}
  \city{Vancouver}
  \country{Canada}
}
\email{wjb@williamjbowman.com}

\begin{abstract}
  Macro embedding is a popular approach to defining extensible shallow embeddings of object languages in Scheme-like host languages.
While macro embedding has even been shown to enable implementing extensible typed languages
in systems like Racket, it comes at a cost: compile-time performance.
In this paper, we revisit \emph{micros}---syntax to intermediate representation (IR) transformers, rather than source syntax to source syntax transformers (macros).
Micro embedding enables stopping at an IR, producing a deep embedding and enabling high performance compile-time functions over an efficient IR, before shallowly embedding the IR back into source syntax.
Combining micros with several design patterns to enable the IR and functions over it to be extensible, we achieve extensible hybrid embedding of statically typed languages with significantly improved compile-time compared to macro-embedding approaches.
We describe our design patterns and propose new abstractions packaging these patterns.
\end{abstract}



\maketitle

{
  \input{intro}
  \input{related}
  \input{extensibility}

  \input{framework}
  \input{eval}

  \input{future}

  \input{acks}
  \input{data-availability}
  \bibliographystyle{ACM-Reference-Format}
  \input{bib.bbl}

}

\end{document}

%% file: intro.tex
\section{Introduction}
Contemporary macro systems have features for linguistic reuse and interposition, enabling macro embedding whole general-purpose languages with sophisticated static type systems as libraries~\cite{tobin-hochstadt2011}.
Macro embedding is even expressive enough to enable user-extensible dependent type systems~\cite{chang2017,chang2020}.

However, macro embeddings can suffer serious performance problems.
Shallowly embedding object languages through macros provides an elaboration into a host language, but means there is never a direct representation of the whole embedded language program.
Exposing such a representation would require stopping macro expansion at some intermediate point and inverting the usual outside-in order of macro expansion, or decompiling fully expanded programs back into an intermediate representation.
This makes implementing optimizations, particularly non-local transformations, difficult or impossible.
Worse, it can dramatically affect compile time.
By inverting expansion order, stopping expansion, or decompiling, the eventually emitted syntax objects will be re-expanded, resulting in worst-case quadratic expansion cost, and thus quadratic time algorithms for what would otherwise be a linear time AST traversal.
As we will show, this is particularly noticeable when embedding typed languages with sophisticated type systems.

An alternative to shallow embedding, deep embedding, would provide a full AST that could be traversed and manipulated efficiently.
Deep embeddings can even be integrated with a macro system, so that the macro system essentially provides the frontend, generating an full AST that can then be manipulated efficiently, before finally shallowly embedding back into the host language~\cite{ee-lib, syntax-spec}.
Unfortunately, this approach gives up on the extensibility of the IR (although the source language remains macro-extensible).

In this paper, we present an approach to hybrid embedding languages with good compile-time performance and extensibility of both the surface language and internal representation. We develop a taxonomy for extensibility, and use it to compare and contrast our approach with the state of the art.
We revisit \emph{micros} as a key abstraction for creating an extensible deeply embedded IR~\cite{krishnamurthi2001}.
Our micros target compile-time structs in Racket for an efficient AST representation.
For extensible manipulation of the object language AST, we use generics to attach each compile-time function (such as the type checker) to its AST node.
We use an extensibility pattern atop structs and generics to navigate the expression problem~\cite{expression}, and propose as future work two new abstractions packaging these patterns.
This enables users to extend the IR both with new nodes and their compile-time functions, and override previous behaviour, similar to Racket objects but with better compile-time performance.
We present a brief case-study of a dependently typed language micro embedded in Racket, extended with gradual typing, and a brief performance analysis of the type checker.

%% file: related.tex
\section{Related Work}
\label{sec:relwork}

To understand the language embedding design space, we first survey existing tools for creating language embeddings, which we refer to as \textit{embedding language frameworks} (\textit{ELFs}).

\paragraph{Turnstile} \texttt{turnstile+} \cite{tsplus}, a dependent-type-supporting extension of the \texttt{turnstile} ELF described by \citet{ts}, uses Racket macros to provide object language extensibility.
\texttt{turnstile+} encodes object language forms as macros and performs type-checking operations as object languages are macro-expanded into core Racket forms.
Adding new forms is as simple as importing a base language and defining new macros for new language forms.
Since Racket's macros expander is open recursive,
redefinition of old language forms is almost as easy: elide old terms in the import specification and define new terms with the same syntax.
Interoperability of object languages with Racket is also easy, since everything expands into a shallow Racket embedding.

While \texttt{turnstile+} offers good extensibility characteristics, it fails to offer good compile time performance (see \Cref{sec:performance}).
Since \texttt{turnstile+} performs type checking during macro expansion, the competing expansion orders of macro expansion and type checking cause a worst-case quadratic expansion time.
Typically, macros are expanded in order from outside in, meaning that in a run-of-the-mill macro invocation a term is walked only once by Racket's macro expander.
Type checking requires sub-terms be checked, and thus expanded, prior to making a judgement on the type of parent terms.
Since sub-terms must be fully expanded into Racket core syntax to be type checked, each term must first be completely expanded, with each sub-term recursively type checked, prior to its shallow embedding being walked by Racket's macro expander \textit{again} after expansion.

In dependent type systems, like those targeted by \linebreak \texttt{turnstile+}, type checking is significantly more complex than for simply typed languages, worsening performance. Furthermore, type checking in \texttt{turnstile+} is performed on syntax objects, requiring the traversal and pattern matching of an inefficient linked-list representation of the object language AST. This incurs poor compile time memory allocation and access performance when compared to a struct-based AST implementation, and makes common optimizations like in-place mutation impossible.

\paragraph{syntax-spec and ee-lib} \texttt{syntax-spec} \cite{syntax-spec} is an ELF which provides a DSL for defining macro-extensible object language syntax, and generates its expansion into an object-language-specific base language. \texttt{syntax-spec} is implemented in \linebreak Racket, and is built over \texttt{ee-lib} \cite{ee-lib}, a compatibility layer between object languages and a host language’s macro and binding system. Languages defined with \texttt{syntax-spec} expand into a base object language without performing type checking or fully elaborating into core Racket syntax. \linebreak \texttt{syntax-spec} does not allow for extension of this base language, instead opting for macro-extensibility of the object language. Once elaborated into the base language, a program written in the object language can then be compiled by any method of the implementer wishes, even one which is non-extensible. \texttt{syntax-spec} also allows an object language to define boundary macros, with which object language users can mix object and host language code.

Given the flexibility it affords in compiler design for the base language, \texttt{syntax-spec} offers good compile time performance; an object language implementer can use whatever performant compiler design they wish for the base language, unconstrained by the \texttt{syntax-spec} parsing and macro frontend. If one were to type-check an object language defined using \texttt{syntax-spec}, a type-checker over the base syntax could be specified as part of the base language compiler, but the \texttt{syntax-spec} framework does not provide any means to extend that type-checker. This is in contrast to \texttt{turnstile+}, which exposes type-checking semantics as an extensible part of the object language.

\paragraph{LMS} Lightweight Modular Staging \cite{lms} (\textit{LMS} for short) is an ELF that directly exposes an object language as a host language data structure, leveraging object-oriented design to achieve extensibility. LMS is implemented as a Scala library, providing the \texttt{Rep[\_]} type for wrapping staged object language forms, allowing composition and manipulation of object language programs using whatever means the language implementer finds appropriate. LMS does not separately define concrete syntax; instead the object language syntax is that of the host language structure---a language user directly writes the abstract syntax representation of their object language program.

Consequentially, LMS encoded languages have no concrete syntax specification. A user of LMS can separately specify concrete syntax using Scala's macro system, which in the context of LMS would function like \texttt{syntax-spec}'s macro frontend, requiring elaboration into an LMS-encoded object language. Unlike \texttt{syntax-spec}, this core language could be extended via LMS.

%% file: extensibility.tex
\section{Extensibility and Flexibility}
\label{sec:extdef}

While intuition reveals some differences in the ELFs we've described, we must develop a taxonomy for describing extensibility to specify our goals for our hybrid micro embedding. We begin by defining notions of syntax and semantics, consider embedding styles for ELFs, and then classify the ELFs we've discussed by their support for the extension of object languages in each of these three dimensions.

\subsection{A Taxonomy for Extensibility}
\label{sec:extclasses}

To describe classes of extensions over languages, we must first be able to describe languages.

Concrete syntax specifications dictate which strings are valid program syntax, declaring what users must write to interact with the features, binding forms, and lexical structures of a programming language. Abstract syntax specifications then declare the structure and organization of program syntax. We think of syntax as a first filter: semantic specifications can be used to reason about any well-formed piece of abstract syntax, but not every well-formed piece of syntax is a valid program.

Our notions of extensibility apply to any means of defining abstract syntax specifications, but in our examples we use the abstractions of \textit{syntax classes} and \textit{meta-variables}. Take for example the abstract syntax definition for the $\lambda$-calculus in \Cref{fig:lc}. We group syntax using the meta-variables $x$ and $t$, allowing us to refer to classes of valid abstract syntax representing variables and expressions. Meta-variables allow syntactic forms to be parameterized on all possible instances of syntax of a particular class; for example the $t$ in the syntax specification $t \ t$ denotes any possible expression. We use \textit{term} to refer to any piece of abstract syntax which abides by a given object language's syntax specification.

\begin{figure}[H]
  \center
  \begin{minipage}{0.45\textwidth}
    \begin{mathpar}
      \begin{array}{rl}
          x &\in \textit{Variable} \\
          t &::= x \mid \lambda x . t \mid t \; t
      \end{array}
    \end{mathpar}
  \end{minipage}\hfill
  \begin{minipage}{0.45\textwidth}
    \judgshape{t \to t'}{}
    \begin{mathpar}
      \text{\defrule*{beta-red-lc}{$\beta$-red} \quad}
      (\lambda x . t) \ t' \to t[x \mapsto t']
    \end{mathpar}
  \end{minipage}\hfill
  \caption{The $\lambda$-calculus with (small-step) reduction semantics, omitting $\alpha$, $\eta$, and substitution.}
  \label{fig:lc}
\end{figure}

Semantic specifications are imposed on terms, encoding properties of a language like which pieces of syntax are well-formed programs, and how well-formed programs behave when they’re executed. Models of language semantics may be abstractly thought of as \textit{judgements} over terms, where the inclusion of syntax in a judgement encodes some meaning. When discussing extensibility, we use \textit{judgement} to refer to all statements one could make over about abstract syntax, including specifications of both static and run-time semantics. For example, consider \Cref{fig:lc}, with the judgement $t \to t'$ expressing the reduction of terms in the $\lambda$-calculus. $t \to t'$ relates (possibly distinct) syntaxes representing terms, where $t$ reduces to the term $t'$ iff $t \to t'$ holds for the pair of syntaxes $(t,t')$. By defining $t \to t'$ for all terms, we encode the semantics of ``reducing a term;'' a term $t$ reduces to $t'$ because $t \to t'$ holds for the pair of terms, and we check if an arbitrary term reduces to another by proving that the pair of terms satisfy the judgement.

\

\textit{Syntactic extension} is the modification of the specifications for the syntactic forms and classes that could potentially make up a program. Syntactic extension is broadly useful: in order to add language features, we would like to be able to add new syntactic forms exposing them. But, \textit{how} might we want to change base language syntax? Is an extension strictly additive, adding classes and cases but not modifying existing syntax, or might an extension have the need to modify existing syntax?

We call extensions which only add syntax \textit{additive} and extensions which change existing syntax \textit{strong}. Note that strong syntactic extensibility need not be strictly more desirable than additive syntactic extensibility, as depending on the goals of an ELF, it may not be practical to provide desired performance or usability characteristics while supporting strong extension. We perform an additive syntactic extension in \Cref{fig:scheme} to extend the $\lambda$-calculus into a Scheme-like well-scoped calculus we'll call $\lambda$-s. The syntactic portion of this extension is the addition of a binding context $\Gamma$, which may either be the empty set $\emptyset$ or a pair of some context and a variable name.

\begin{figure}[H]
  \center
  \begin{minipage}{0.20\textwidth}
    \begin{mathpar}
      \begin{array}{rl}
          x &\in \textit{Variable} \\
          t &::= x \mid t \; t
      \end{array}
    \end{mathpar}
  \end{minipage}\hfill
  \begin{minipage}{0.20\textwidth}
    \begin{mathpar}
      \begin{array}{rl}
          \textcolor{blue}{\Gamma} &\textcolor{blue}{::= \emptyset \mid \Gamma, x}
      \end{array}
    \end{mathpar}
  \end{minipage}\hfill
  \vspace{0.5em}
  \judgshape{t \to t'}{}
  \begin{mathpar}
    \text{\defrule*{beta-red-lc}{$\beta$-red} \quad}
    (\lambda x . t) \ t' \to t[x \mapsto t']
  \end{mathpar}
  \vspace{-0.5em}
  \color{blue}
  \judgshape{\Gamma \vdash t}{}
  \begin{mathpar}
    \inferrule[\defrule{Gamma}]{
        x \in \Gamma
      }
      {
        \Gamma \vdash x
      }

      \inferrule[\defrule{Lambda}]{
        \Gamma, x \vdash t
      }
      {
        \Gamma \vdash \lambda x . t
      }

      \inferrule[\defrule{App}]{
        \Gamma \vdash t \\
        \Gamma \vdash t'
      }
      {
        \Gamma \vdash t \ t'
      }
  \end{mathpar}
  \color{black}
  \caption{Extension of the $\lambda$-calculus with a well-scopedness judgement, forming $\lambda$-s. \textcolor{blue}{Additive extensions in blue.}}
  \label{fig:scheme}
\end{figure}

\textit{Semantic extension} provides new meaning to terms. Consider again the $\lambda$-s extension in \Cref{fig:scheme}. To complete the new language feature, we need to add the judgement $\Gamma \vdash t$ to encode the semantics of a "term being well-scoped." We call the introduction of $\Gamma \vdash t$ a \textit{judgement-level semantic extension}, which is the means by which we add new judgements or modify the dependencies between existing judgements. Analogous to syntactic extension, we introduce the notions of additive and strong judgement-level semantic extensibility. Here, we need only additive judgement-level semantic extensibility since we did not need to change any existing judgements, namely $t \to t'$.

Suppose we want to introduce types to this well-scopedness judgement, producing a simply-typed $\lambda$-calculus (\textit{STLC}), with Church-style intrinsic typing. We perform both kinds of syntactic extension in \Cref{fig:stlc}, using additive syntactic extension to add a meta-variable for types, $\tau$, and syntax for function types, constants, and base types. We then strongly extend the existing syntax for terms $t$, modifying the syntax for functions $\lambda x . t$ to include a type annotation, and do similarly for contexts $\Gamma$:

\begin{figure}[H]
  \center
    \begin{minipage}{0.15\textwidth}
    \begin{mathpar}
      \begin{array}{rl}
          x &\in \textit{Variable} \\
          \textcolor{blue}{T} &\textcolor{blue}{\in \textit{Base Types}}\\
          \textcolor{blue}{c} &\textcolor{blue}{\in \textit{Constants}}\\
      \end{array}
    \end{mathpar}
  \end{minipage}\hfill
  \begin{minipage}{0.25\textwidth}
    \begin{mathpar}
      \begin{array}{rl}
          t &::= x \mid \hcancel[red]{\lambda x . t} \enspace \textcolor{red}{\lambda x : \tau . t} \mid t \; t \ \textcolor{blue}{\mid c} \\
          \textcolor{blue}{\tau} &\textcolor{blue}{::= \tau \to \tau \mid T}\\
          ~\\
          \Gamma & ::= \emptyset \mid \hcancel[red]{\Gamma, x} \enspace \textcolor{red}{\Gamma, x : \tau}
      \end{array}
    \end{mathpar}
  \end{minipage}\hfill
  \judgshape{t \to t'}{}
  \begin{mathpar}
    \text{\defrule*{beta-red}{\textcolor{red}{$\beta$-red}} \quad}
    \hcancel[red]{(\lambda x . t) \ t'} \enspace \textcolor{red}{(\lambda x : \tau . t) \ t'} \to t[x \mapsto  t']
  \end{mathpar}~\\
  \judgshape{\Gamma \vdash t \ \textcolor{red}{ : \tau}}{}
  \begin{mathpar}
    \inferrule[\defrule*{gamma-stlc}{\textcolor{red}{Gamma}}]{
        x \ \textcolor{red}{: \tau}\in \Gamma
      }
      {
        \Gamma \vdash x \color{red}{: \tau}
      }

      \color{blue}
      \inferrule[\defrule*{const-stlc}{Const}]{
        c \text{ is of type } T
      }
      {
        \Gamma \vdash c : T
      }
      \color{black}

      \inferrule[\defrule*{lambda-stlc}{\textcolor{red}{Lambda}}]{
        \Gamma, x \ \textcolor{red}{: \tau} \vdash t \ \textcolor{red}{: \tau}
      }
      {
        \Gamma \vdash \lambda x \ \textcolor{red}{: \tau} . t \ \textcolor{red}{: \tau \to  \tau'}
      }

      \inferrule[\defrule*{app-stlc}{\textcolor{red}{App}}]{
        \Gamma \vdash t  \ \textcolor{red}{: \tau \to \tau'} \\
        \Gamma \vdash t' \ \textcolor{red}{: \tau}
      }
      {
        \Gamma \vdash t \ t' \ \textcolor{red}{: \tau'}
      }
  \end{mathpar}
  \caption{An extension of $\lambda$-s into STLC. \textcolor{blue}{Additive extensions in blue} and \textcolor{red}{strong in red}.}
  \label{fig:stlc}
\end{figure}

In our STLC example, we use strong judgement-level extension to add the $\Gamma \vdash t : \tau$ judgement, which modifies the scopedness judgement $\Gamma \vdash t$. Judgement-level extension complete, we have a new class of statements we can make about terms, but have not yet described any such statement. To do so, we must also perform a \textit{rule-level semantic extension,} where we extend meaning to new terms under existing judgements. Rule-level extensions can also be either additive or strong. Here, we use additive rule-level semantic extensibility to add \refrule*{const-stlc}{Const}, which types constants.
We then use strong rule-level extension to modify the former $\Gamma \vdash t$ rules, resulting in typing rules which respect our new $\Gamma \vdash t : \tau$ judgement. As it stands, no other judgements in our STLC specification make use of the $\Gamma \vdash t : \tau$ judgement, but it remains exposed to a user of the ELF as a judgement one can use to describe STLC programs.

Note that additive syntactic extensions may require strong semantic extensions in order to achieve desired behavior. For example, sound gradual typing is a typing discipline which allows for static imprecision in typing that induces run-time constraint checks \cite{gradualtyping}. Often presented in literature as a modification of some base type system, the run-time semantics of a gradually typed language differ from those of the base language, changing how existing terms are interpreted. To add gradual types to a simply typed system, we would need to add the unknown type, an additive syntactic extension. Semantic extensions supporting this new syntax must be made to existing judgements, strongly extending rule-level semantics. We summarize our extensibility classes in \Cref{fig:classes}.

\begin{table*}[h]
\caption{Extensibility classes.}
\label{fig:classes}
\centering
\begin{tabular}{ |c||c|c|c| }
  \hline
  \multirow{2}{*}{\backslashbox{\textbf{Strength}}{\textbf{Domain}}} & \multirow{2}{*}{\textbf{Syntactic}} & \multicolumn{2}{c|}{\textbf{Semantic}} \\
  \cline{3-4}
  & & \textbf{Rule-level} & \textbf{Judgement-level} \\
  \hline
  \hline
  \textbf{Additive} & \makecell{Addition of \\ syntactic forms} & \makecell{Addition of syntax to \\ existing judgements} & \makecell{Addition of new \\ judgements} \\
  \hline
  \textbf{Strong} & \makecell{+ modification of \\ existing syntactic forms} & \makecell{+ modification of judgements \\ over existing syntax} & \makecell{+ modification of \\ existing judgements}\\
  \hline
 \end{tabular}
\end{table*}

The classes of extensibility we define map well to the \textit{expression problem}, which as described by \citet{expression}, is the problem of achieving extensibility of both data and addition of functions which operate on that data. In the context of programming languages, terms are data, and language semantics are functions over terms. Designing ELFs that support both rule-level and judgement-level semantic extension requires one to solve the expression problem. We require extensibility of data to achieve syntactic and rule-level semantic extension, and extensibility of functions on that data for judgement-level semantic extension. To support \textit{either} rule-level semantic extension or judgement-level semantic extension, one does not need to solve the expression problem. Rule-level extension is modelled with object-oriented programming (\textit{OOP}) structures, where syntax maps to classes, and judgements to methods called by a visitor, but adding new methods is hard. In contrast, judgement-level extension maps to typical functional patterns, where writing new functions which operate on an existing datatype is easy, but adding new cases to union types requires modification of existing functions.

Lastly, we define \textit{extension} and \textit{modification}. In Wadler’s presentation, a solution to the expression problem should allow for extension without modification and re-compilation of (in our context) a base language's compiler; if the original language no longer exists as a distinct entity---a distinct unit of compilation in the host language---the object language has not been extended, it has been modified.

\subsection{A Taxonomy for Flexibility}
\label{sec:flexibility}

An embedding ELF's choice of embedding strategy greatly impacts the flexibility of the object language, be it \textit{deep}, \textit{shallow}, \textit{hybrid} \cite{foldds}, or something less common like \textit{compositional} \cite{comp} or \textit{polymorphic} \cite{poly} embeddings. The related works discussed in this paper implement some variation of the first three, summarized below:

\begin{itemize}
  \item \textbf{Deep embeddings} encode object languages as an AST which can be traversed and manipulated in the host language. Whole-program transformations can traverse the entire AST of object language programs, and a final step in compilation transforms the object language AST into host language code. Since the encoding of object language terms is different from those of equivalent host language terms, deep embeddings leak implementation details which prevent naive interoperability with equivalent host language features.
  \item \textbf{Shallow embeddings} encode object languages directly in terms of equivalent host language functionality. A shallowly embedding ELF can never reason about an AST as a whole, and as a consequence cannot perform whole-program transformations. However, shallow embeddings benefit from much better host language interoperability, as host language features can be used directly on encoded object language terms, as their encoding is idiomatic in the host language.
  \item \textbf{Hybrid embeddings} construct an AST as a deep embedding would, upon which whole-program transformations can be defined, but allows host language features to interact with object language terms like a shallow embedding. Hybrid embedded object language terms have both an internal representation (used to construct a whole-program AST) and a wrapper allowing for direct host language interoperability.

\end{itemize}

We note that with shallow embeddings, host and object language interoperability generally is possible at term boundaries; one can pass object language terms to host language functions and vice versa. With hybrid embeddings, an object language author must choose at which granularity to force the construction of a complete object language AST. In order to perform meaningful optimizations, it may be the case that interaction between host and object language can only occur at, for example, module boundaries.

\subsection{Related Work, Taxonomized}

Having established our taxonomy, we re-examine the ELFs discussed in \Cref{sec:relwork}.

\paragraph{Turnstile} Recall that \texttt{turnstile+} encodes language \linebreak forms as host-language macros, which can be imported from a base language and shadowed by language extensions.
Since each macro is responsible for the transformation and type-checking of its associated syntactic form in the object language, macro shadowing allows \texttt{turnstile+} to support \linebreak strong syntactic and strong rule-level semantic extension.
\linebreak \texttt{turnstile+} also supports strong judgement-level semantic extension, as new terms can implement new judgements, and by shadowing old terms, old judgements can be modified for existing terms. Given \texttt{turnstile+} language forms are Racket macros, which individually expand fully to Racket core forms, \texttt{turnstile+} is a shallow embedding ELF. \linebreak \texttt{turnstile+} object language terms can be individually imported and used in (host) language code, and a user does not have access to arbitrary AST transformations.

\paragraph{syntax-spec} A Racket-embedded DSL framework, \linebreak \texttt{syntax-spec} provides a syntax specification language and macro system for object languages, with a conventional compiler design downstream. This approach requires that all extensions must live within an \textit{object language} macro system. This is in contrast to \texttt{turnstile+}, where extensions are written using host-language level macros. Consequentially, one cannot \textit{extend} a \texttt{syntax-spec} encoded language if it turns out that new constructs need to be introduced to the core in order to support a desired feature, but must instead modify the \texttt{syntax-spec} definition for the DSL core in order to express novel semantics.

As an example, in STLC as we defined in \Cref{fig:stlc}, we could not add types previously unknown to the compiler.
In \texttt{syntax-spec}, object-language extensions are limited to using macros to translate new syntax into existing core syntax.
That is, the object language is macro extensible \cite{felleisen1991}, but neither truly syntactically extensible nor truly semantically extensible in our taxonomy, since the core language cannot be extended.
This is in contrast to \texttt{turnstile+}, where the \linebreak \textit{compiler} is specified extensibly, and can be iterated upon without clobbering the compiler for a base language.
\linebreak \texttt{syntax-spec} also allows an object language specification to define macros which introduce an interaction point between host and object language terms, making \texttt{syntax-spec} a hybrid embedding ELF.

\paragraph{LMS} LMS supports strong syntactic extension, and both strong rule-level and strong judgement-level semantic extension. Since LMS object language constructs are encoded as Scala classes, LMS is a deeply embedding ELF. Scala's OOP and trait system allows LMS to achieve these extensibility properties, at the expense of poor host language interoperability. LMS's \texttt{Rep[\_]} type, used to stage object language programs, leaks into instances of object language syntax and typing, making object language representations incompatible with equivalent Scala features. LMS thus deeply embeds its object language.

%% file: framework.tex
\section{A Micro Embedded Framework}
\label{sec:flex}

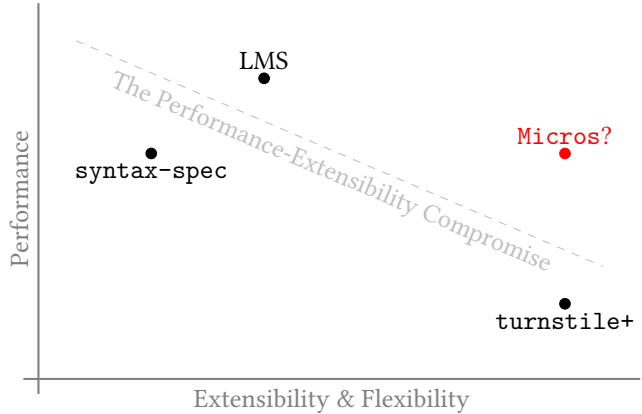
\begin{figure}[h]
\center
\begin{tikzpicture}
  \draw[gray, thick] (-0.2,0) -- (8,0) node[midway,below] {Extensibility \& Flexibility};
  \draw[gray, thick] (0,-0.2) -- (0,5) node[midway,above,rotate=90] {Performance};
  \filldraw[black] (1.5,3) circle (2pt) node[anchor=north]{\texttt{syntax-spec}};
  \filldraw[black] (3,4) circle (2pt) node[anchor=south]{LMS};
  \filldraw[black] (7,1) circle (2pt) node[anchor=north]{\texttt{turnstile+}};
  \filldraw[red] (7,3) circle (2pt) node[anchor=south]{\texttt{Micros}?};
  \draw[lightgray, dashed] (0.5,4.5) - - (7.5,1.5) node[midway,below,rotate=-23] {The Performance-Extensibility Compromise};
\end{tikzpicture}
  \caption{The ELF design space, optimistically.}
\label{fig:domain}
\end{figure}

We want to determine if it is possible for a hybrid embedding ELF to support strong syntactic, strong rule-level and strong judgement-level semantic extensibility, while compiling a dependently typed object language faster than \texttt{turnstile+}. Ideally, we want compilation performance on par with an object language implemented with an unrestricted choice of compiler architecture. To evaluate this research question, we begin developing a Racket-hosted ELF, using micros as a syntactic frontend which elaborates into a struct-based IR, made extensible using Racket generics. Given our analysis of related work, we can crudely draw the ELF design space in \Cref{fig:domain}. Note the gap between \texttt{turnstile+} and \texttt{syntax-spec}/LMS; this is the gap between the extremes of extensibility and performance we want to target with our micro embedding framework.

We present our micro embedding strategy as a collection of programming patterns, which we realize in Racket.
To evaluate the extensibility of micro embeddings, we implement \texttt{fowl-base}, a bidirectional Martin-Löf type theory with natural, boolean, equality and $\Pi$ types, which we extend independently with vector and $\Sigma$ types. We then inherit from both extensions to form \texttt{fowl}, which we use as a base to further interrogate the extensibility properties of micro embeddings. We extend the \texttt{fowl} type checker with gradual dependent types \cite{geq} to create \texttt{fowl-geq}, and attempt to add exceptional types \cite{rett} to create \texttt{fowl-rett}. Our IR is unable to encode the \texttt{fowl-rett} extension; we discuss future work required to support \texttt{fowl-rett} in \Cref{sec:future}. We conclude in \Cref{sec:evalext} that \texttt{fowl} strongly supports syntactic and judgement-level semantic extension, but only additive rule-level extensibility. In \Cref{sec:performance}, we conclude that compilation of \texttt{fowl} performs two orders of magnitude better than a similar language written in \texttt{turnstile+} when evaluated on a dependently typed benchmark suite.

\subsection{The Problem with Macros}

Consider the following Racket pseudocode, shallowly embedding an object language \texttt{my-if} construct in terms of Racket's native \texttt{if}:

\begin{minted}[fontsize=\footnotesize]{racket}
(define-syntax (my-if stx)
  (syntax-parse stx
    [(_  pred con alt) #'(if pred con alt)]))
\end{minted}

Easy, breezy, beautiful.
But, what happens when we want to transform a program that uses \texttt{my-if}?
Suppose we'd like to perform a static optimization on \texttt{my-if}, compiling \texttt{my-if} to \texttt{con} if \texttt{pred} is a tautology.
A naïve option is to pattern match the syntax of \texttt{pred}, using some internal logic to search for syntactic patterns that will always yield true a run time:

\begin{minted}[fontsize=\footnotesize]{racket}
(define-syntax (my-if stx)
  (syntax-parse stx
    [(_  my-true con alt) #'con]
    [(_  (my-not my-false) con alt) #'con]
    ;; ... and many more
    [(_  pred con alt) #'(if pred con alt)]))
\end{minted}

In general, our macro could thread some context through expansion to aid in symbolic execution, making this approach effective.
However, it is also fragile: any change to program syntax, such as macro extensions of the source language, defeats this optimization.
If we add new core syntax, we need to \emph{modify} \texttt{my-if} to include those cases as well.
This fails to meet our criterion for extensibility.

Suppose then that we require all forms to expand to their optimized core representation, and use \texttt{local-expand} to invert expansion order, expanding sub-expressions first into core forms.
All macro extensions will be elaborated away, and core forms will optimize themselves if possible.
Then, we can implement \texttt{my-if} as follows.

\begin{minted}[fontsize=\footnotesize]{racket}
(define-syntax (my-if stx)
  (syntax-parse (local-expand stx)
    [(_  #t con alt) #'con]
    ;; ... and many more
    [(_  pred con alt) #'(if pred con alt)]))
\end{minted}

Now, we've rediscovered the implementation strategy used by \texttt{turnstile+}!
In \texttt{turnstile+}, sub-terms are type checked and manipulated by invoking \texttt{local-expand}, allowing extensions to be transparent to existing macros.
For tasks like normalization, \texttt{turnstile+} can parse Racket core forms as the core IR.

Unfortunately, while extensible, this approach causes performance problems.
Using \texttt{local-expand} and then including the emitted syntax object in our macro's output can cause quadratic expansion times, since the emitted syntax object (and all of its sub-expressions) will be re-expanded.
For computationally intensive passes, such as dependent type checking algorithms, relying on syntax objects (an immutable linked list data structure) hampers performance optimizations we might want to perform.

All of these problems stem from macros expanding into syntax objects, and staying in the macro expander.
So why not elaborate into something else, and interrupt macro \linebreak expansion?

To implement this approach, we can use \textit{micros} \cite{krishnamurthi2001}, which are transformers from syntax to a core IR, rather than syntax to syntax.
Each micro expands into a core IR term, represented as an arbitrary data structure, and no more macro expansion happens on it.
For example, we might define a \texttt{my-if} \emph{micro} using structs as our IR representation.

\begin{minted}[fontsize=\footnotesize]{racket}
(begin-for-syntax
  (struct if-ast (pred con alt)))
(define-syntax (my-if stx)
  (syntax-parse stx
    [(_  pred con alt)
     (if-ast (local-expand #'pred)
             (local-expand #'con)
             (local-expand #'alt))]))
\end{minted}
This yields a deep embedding.
If we use (compile-time) structs with generic methods attached, we can expand into an extensible core IR with many additional extensible compiler passes, such as type checking, before eventually shallow embedding back into syntax objects.
This micro cannot be used by Racket's macro expander directly, since the macro expander expects transformers to produce syntax objects. We will need to trick the expander into letting us produce other data structures.

In the rest of this section, we present using this approach to implement \texttt{fowl-base}, an extensible dependently typed language hybrid embedded into Racket.

\subsection{fowl-base-syn}

\texttt{fowl-base} is two Racket modules. The first, \texttt{fowl-base-syn}, uses micros to define the syntax of \texttt{fowl-base} and its elaboration into an IR defined in the module \texttt{fowl-base-sem}.

Like \texttt{turnstile+} and \texttt{syntax-spec}, \texttt{fowl-base}'s micros are written using Racket macros.
\texttt{fowl-base-syn} exports all defined micros, which can be imported either individually for term-level interoperability, or as a Racket hashlang in which a user can write \texttt{fowl-base} language programs.
\Cref{fig:syndef} includes the micro definition for \texttt{if}, which expands into an abstract syntax representation.

Racket's module system makes it easy to extend our micro embeddings: an extension imports the syntax and semantics modules from a base language, and exports its own language forms, shadowing the base language. To perform a weak syntactic extension to \texttt{fowl-base}, we define a new micro in our extension. If we desire strong extension, this new micro can shadow the name of the old micro we wish to override. Modules also allow for separate compilation of a base and extended language.

\begin{figure}[h]
\begin{minted}[fontsize=\footnotesize]{racket}
(define eeyore void)

(begin-for-syntax
  (define mule #'eeyore)
  (define (burden-mule expansion)
    (syntax-property mule 'expansion expansion))
  (define (unburden-mule m)
    (syntax-property m 'expansion))

  (define (elab-to-structs e)
    (define idx (syntax-local-make-definition-context))
    (syntax-parse (local-expand e 'expression '() idx)
      [e:expanded-term (unburden-mule #'e.body)])))

(define-syntax (if stx)
  (syntax-parse stx
    [(_  pred con alt)
     (burden-mule (fi:if-term (elab-to-structs #'pred)
                              (elab-to-structs #'con)
                          (elab-to-structs #'alt)))]))
  \end{minted}
    \caption{An excerpt of \texttt{fowl-base-syn} demonstrating the \texttt{mule} micro pattern.}
  \label{fig:syndef}
  \vspace{-1em}
\end{figure}

We want micros to expand into structs, but Racket macros are syntax to syntax transformers; we need a way to return data instead.
We implement micros on top of macros using the \textit{mule pattern} demonstrated in \Cref{fig:syndef}.
The \emph{mule pattern} creates a dummy piece of syntax, \texttt{\#'eeyore}, and attaches to it the micro-expansion of the term as metadata (using syntax properties).
The dummy syntax requires no further expansion, avoiding re-expansion costs.
Our micros all follow this pattern, using a macro to define surface syntax, and the mule pattern to compose and return our IR representation, carrying it through macro expansion.
\texttt{elab-to-structs} is used to recursively expand micros, returning the struct representations of child terms.

\subsection{fowl-base-sem}

\texttt{fowl-base-sem} defines mutable \textit{structures}, one for each \linebreak \texttt{fowl-base} language term.
Racket structures are named records, which can optionally support \textit{generic methods} (or in OOP nomenclature, interface methods) that can be called on any instance of a structure implementing the generic.
Using structures as our IR representation improves performance over a syntax object representation, as structure fields have constant time access, and mutability allows the IR to be manipulated in place by language transformations.

\texttt{fowl-base-sem} also defines rules and judgments over the struct IR.
To maintain extensibility, each \texttt{fowl-base} judgement is defined as a generic interface.
To add a judgement rule for a \texttt{fowl-base} term, we attach an implementation for the judgement's generic interface to the structure in \texttt{fowl-base-sem} representing the \texttt{fowl-base} term.

Structs in an extension can implement a generic method for a judgement related to their AST node, giving us additive rule-level semantic extension.
Note that this requires that judgement rules are syntax-directed.
Additive judgement-level extensibility is also easy: extensions can define a new generic interface for new judgements, using \texttt{\#:defaults} to implement rules for base language structs.

We see generics as judgements in practice in \Cref{fig:termdef}.
\texttt{fowl-base} is a bi-directionally typed language, so terms may implement either a type synthesis judgement, a type checking judgement, or both.
In \Cref{fig:termdef} we define the synthesis and check judgements of \texttt{fowl-base} and a portion of the internal representation of \texttt{suc}, the successor constructor for Peano numerals.
Here, we implement the inference rule for the synthesis judgement.
The statement \texttt{\#:methods gen:ir-elaborable} denotes that \texttt{suc} implements the generic \texttt{ir-elaborable} interface, and thus the \texttt{elab-ir-synth!} and \texttt{elab-ir-check!} judgements.
The choice to define both \texttt{synth!} and \texttt{check!} under the same interface is one of convenience, since in \texttt{fowl-base} all terms happen to implement both judgements.
Unlike \texttt{turnstile+}, the synthesis judgement for \texttt{fowl-base} is able to mutate structures in place, meaning that when \texttt{fowl-base-syn} expands \texttt{fowl-base} surface syntax into \texttt{fowl-base-sem} structures, the synthesis and check judgements do not allocate and return a new syntax tree, but instead manipulate the input abstract syntax in place.

\begin{figure}[h]
  \begin{minted}[fontsize=\footnotesize]{racket}
    (define-generics ir-elaborable
      [elab-ir-synth! ir-elaborable t-env v-env r-env]
      [elab-ir-check! ir-elaborable t t-env v-env r-env])

    (parameterize-judgement elab-ir-synth!)
    (parameterize-judgement elab-ir-check!)

    (serializable-struct
      suc term (nat) #:mutable #:transparent
      #:methods gen:ir-elaborable
      [(define (elab-ir-synth! e t-env v-env r-env)
          (match-define (suc n) e)
          (let ([n~ (elab-ir-check!$ n Nat
                                     t-env v-env r-env)])
            (when n~ (set-suc-nat! e n~))
            (values #f Nat)))])
  \end{minted}
  \caption{An excerpt of \texttt{fowl-base-sem} demonstrating a judgement and term definition.}
  \label{fig:termdef}
\end{figure}

While the patterns we've covered thusfar suffice to provide additive extensibility, they do not provide \textit{strong} semantic extensibility.
For strong extensibility, we need some way to dynamically bind rules and judgements, so that we can override base language behavior with new behavior defined in an extension.
To achieve this, we introduce interposition points for judgements using Racket parameters, which can be dynamically rebound.
We call this the \emph{extensible generics} pattern.

The extensible generics pattern introduces a parameter for each generic, which acts as an interposition point for the judgement.
All extensible generics call their implementation through this parameter, so by modifying the parameter, the judgement can be strongly extended.

In \Cref{fig:termdef} we call \texttt{parameterize-judgement}, a helper macro that uses inserts a parameter for \texttt{fowl-base} judgements.
We present the full definition in \Cref{fig:parjudg}.
Given the name of a judgement's generic function \texttt{A}, the macro wraps \texttt{A} in a parameter using \texttt{make-parameter}, and binds it to \texttt{A-prm}.
\texttt{parameterize-judgement} also creates a memoized accessor procedure for this parameter, named \texttt{A\$}.
We then use \texttt{A\$} in the rest of \texttt{fowl-base} to invoke the judgement encoded by our original function \texttt{A}.
Memoization is used to improve compile time performance, as parameter bindings do not change as a micro embedded language compiler is running, but are accessed regularly.

\begin{figure}[h]
  \begin{minted}[fontsize=\footnotesize]{racket}
    (define-syntax (parameterize-judgement stx)
    (syntax-parse stx
      [(_ gen)
       (let ([name (format-id #'gen "~a-prm" #'gen)]
             [acc  (format-id #'gen "~a$"    #'gen)])
         #`(begin (define #,name (make-parameter gen))
                  (define #,acc (memo-prm #,name))))]))
  \end{minted}
    \caption{\texttt{parameterize-judgement} macro for parametric binding of object language judgements.}
  \label{fig:parjudg}
  \end{figure}

Introducing an interposition point for judgement forms gives us strong judgement-level semantic extensibility.
\linebreak Thanks to these interposition points, languages extending \texttt{fowl-base} can modify judgements over \texttt{fowl-base} terms without requiring the recompilation of \texttt{fowl-base-sem}.
Since \texttt{A-prm} is dynamically bound, we can redefine it in a language extension, resulting in all instances of \texttt{A\$} within scope of that redefinition (including those of the base language) changing.
We give an example of this with \texttt{fowl-geq} in \Cref{sec:evalgeq}.

Not shown is a pattern for obtaining strong \textit{rule}-level semantic extensibility. To do so, we would require interposition points for structs, allowing us to rebind constructors for base language structs. Unfortunately, this was only realized during the implementation of \texttt{fowl-rett}, and thus \texttt{fowl-base} achieves only weak rule-level semantic extensibility. \Cref{sec:evalrett} expands on the consequences of this omission, and we consider approaches for fixing this in \Cref{sec:future}.
%

%% file: eval.tex
\section{Evaluation}
\label{sec:eval}

To evaluate the extensibility of our micro embedding approach, we extend \texttt{fowl}'s type system.
To quantify the performance of micro embeddings, we compile a suite of dependently typed programs in both \texttt{fowl} and \texttt{Cur}, a dependently typed calculi implemented in \texttt{turnstile+}.

\begin{figure}[H]
  \center
  \begin{tikzpicture}
    \draw[gray, dashed, -{Stealth[length=4mm, open, round]}] (-2,0) -- (0,1);
    \draw[gray, dashed, -{Stealth[length=4mm, open, round]}] (-2,0) -- (0,-1);
    \draw[gray, dashed, -{Stealth[length=4mm, open, round]}] (0,1) -- (2,0);
    \draw[gray, dashed, -{Stealth[length=4mm, open, round]}] (0,-1) -- (2,0);
    \filldraw[black] (0,1) circle (2pt) node[anchor=south]{\texttt{fowl-vec}};
    \filldraw[black] (0,-1) circle (2pt) node[anchor=north]{\texttt{fowl-sigma}};
    \filldraw[black] (-2,0) circle (2pt) node[anchor=east]{\texttt{fowl-base}};
    \filldraw[black] (2,0) circle (2pt) node[anchor=west]{\texttt{fowl}};
  \end{tikzpicture}
  \caption{\texttt{fowl}'s diamond inheritance structure.}
  \label{fig:diamond}
\end{figure}
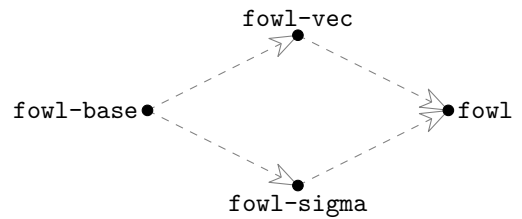

\subsection{Extensibility}
\label{sec:evalext}

\subsubsection{Additive Extension}

To test additive extensibility, we define two extensions to \texttt{fowl-base}, \texttt{fowl-vec} and \linebreak \texttt{fowl-sigma}, and combine them to create \texttt{fowl}; \Cref{fig:diamond} summarizes this inheritance pattern. In \Cref{fig:fowlvecdefs} we include an excerpt of \texttt{fowl-vec}'s implementation. Here, \texttt{fowl-vec} imports \texttt{fowl-base} and defines new macros and structures for each of \texttt{fowl-vec}'s new terms, exporting both the new definitions and those original to \texttt{fowl-base}. \texttt{fowl-sigma} is written similarly, and is omitted here. Since we never needed to modify \texttt{fowl-base}, it remains its own language, and can be used as before in \texttt{fowl-base} language programs.

\begin{figure}
\begin{minted}[fontsize=\footnotesize]{racket}
;; fowl-vec-sem
(require fowl-base-sem)
(provide (all-defined-out)
         (all-from-out fowl-base-sem))

(serializable-struct vec-term ground (len type) ...)
\end{minted}
\begin{minted}[fontsize=\footnotesize]{racket}
;; fowl-vec-syn
(require fowl-base-syn (for-syntax fowl-vec-sem))
(provide (all-from-out fowl-base-syn)
         Vec nil :: ind-Vec)

(define-syntax (Vec stx)
  (syntax-parse stx
    [(_ len type)
     (burden-mule (vec-term (elab-to-structs #'len)
                            (elab-to-structs #'type)))]))
  \end{minted}
  \caption{An excerpt of \texttt{fowl-vec-syn} and \texttt{-sem}.}
  \label{fig:fowlvecdefs}
\end{figure}

\texttt{fowl} can then be defined by the code in \Cref{fig:fowldefs}. Since both \texttt{fowl-vec} and \texttt{fowl-sigma} make additive, nonconflicting extensions, all that is required is a simple dependency merge.
We conclude that \texttt{fowl} supports at least additive syntactic extension and additive rule-level semantic extension.

\begin{figure}[h]
  \begin{minted}[fontsize=\footnotesize]{racket}
    ;; fowl-syn
    (require fowl-vec-sem fowl-sigma-sem)
    (provide (all-from-out fowl-vec-sem fowl-sigma-sem))
  \end{minted}
  \begin{minted}[fontsize=\footnotesize]{racket}
    ;; fowl-sem
    (require fowl-vec-syn fowl-sigma-syn)
    (provide (all-from-out fowl-vec-syn fowl-sigma-syn))
  \end{minted}
  \caption{The entirety of \texttt{fowl-syn} and \texttt{fowl-sem}.}
  \label{fig:fowldefs}
\end{figure}

\subsubsection{Judgement-level Semantic Extension}
\label{sec:evalgeq}

To test \linebreak judgement-level semantic extensibility, we extend \texttt{fowl} with gradual types.
\textit{GEq} \cite{geq} is a dependently typed language with gradual types and sound equality over gradually typed terms. Relative to Bidirectional-CIC, GEq modifies the semantics and syntax of \texttt{J}, \texttt{Equal} and \texttt{refl}, and introduces a new judgement we name \texttt{elab-geq-coerce-unk!}, that coerces the unknown type to a term with a type requested by a check judgement.

In order for \texttt{fowl-geq} to coerce unknown terms in its check judgement, we use strong judgement-level extensibility to shadow \texttt{fowl}'s old check judgement with the modified one we define in \texttt{fowl-geq-sem}.
\Cref{fig:fowlgeqjudgement} shows how we achieve this with Racket's parameters: When calling into the entry point of \texttt{fowl}'s type checker, we \texttt{parameterize} the call with the function implementing \texttt{fowl-geq}'s instance of the check judgement, rebinding all uses of check in \texttt{fowl}.
We also use strong syntactic extension in order to shadow \texttt{fowl}'s definitions for \texttt{J}, \texttt{Equal} and \texttt{refl}, extending their syntax with that required by GEq.
The new equality terms elaborate into new \texttt{J-geq}, \texttt{equal-geq} and \texttt{refl-geq} structures in \texttt{fowl-geq-sem}.
While normally we'd need to use strong rule-level extensibility to revise the check judgement for the old \texttt{J}, \texttt{Equal} and \texttt{refl} terms, they are never constructed by any other term in \texttt{fowl}, and therefore the old check implementation for these terms will never be invoked by a \texttt{fowl-geq} program.

\begin{figure}[h]
  \begin{minted}[fontsize=\footnotesize]{racket}
(define (elab-geq-constr-synth! e h args
  t-env v-env r-env) #| Implementation ... |# )

(define (elab-geq e)
  (parameterize
    ([elab-ir-constr-synth!-prm elab-geq-constr-synth!]
     [type-consistent?-prm type-consistent-geq?])
    (define-values (e^ e^-t) (elab-ir-synth!$ e (empty-env) (empty-env) (empty-env)))
    (values (if e^ e^ e) e^-t)))
  \end{minted}
  \caption{A snippet of \texttt{fowl-geq-sem} rebinding a judgement form.}
  \label{fig:fowlgeqjudgement}
\end{figure}

\subsubsection{Rule-level Semantic Extension}
\label{sec:evalrett}
To test rule-level semantic extensibility, we extend \texttt{fowl} with exceptions.
Reasonably Exceptional Type Theory \cite{rett}, or \textit{RETT}, is an extension of CIC that adds exceptions that can be thrown and caught. Unlike CIC, which has a single hierarchy of type universes (\texttt{(Type 0)} being the smallest of such universes in \texttt{fowl-base}, represented by the \texttt{sort-term} structure type in \texttt{fowl-base-sem}), RETT has three separate universe hierarchies. As a consequence, \texttt{sort-term} now needs to store an additional field: not only the universe level, but also which hierarchy that universe belongs to; let's call its replacement \texttt{sort-term-rett}. In this case, only some of \texttt{fowl}'s terms need to interact with \texttt{sort-term-rett} any differently than they would with \texttt{sort-term}.

Unfortunately, the evaluated version of \texttt{fowl} does not have interposition points for abstract syntax constructors. Consequentially, we cannot replace all instances of the \linebreak \texttt{sort-term} constructor in \texttt{fowl} with \texttt{sort-term-rett}, defined in \texttt{fowl-rett}, without replacing the judgements that refer to \texttt{sort-term}. To deal with this, we have two options:

\begin{enumerate}
    \item Replace all terms that interact with \texttt{sort-term} with new \texttt{fowl-rett} terms, as we did with \texttt{fowl-geq}.
    \item Create new check and synthesis judgements, taking advantage of strong judgement-level extensibility, and replace the judgements in \texttt{fowl} with these new judgements, which support \texttt{sort-term-rett}.
\end{enumerate}

While both options would \textit{work}, we hesitate to legitimize either as an extension; we'd be replacing a large portion of either \texttt{fowl} terms, or the synthesis and check judgements in \texttt{fowl}, that have nothing to do with the change we actually want to make.
Besides being annoying, it's bad programming practice---changes and fixes made to \texttt{fowl} will not propagate to \texttt{fowl-rett}.
To fix this, we need interposition points on struct constructors, but this is nontrivial to support with generics; we discuss this further in~\Cref{sec:future}.

\subsubsection{Micro Embeddings, Taxonomized}
We conclude the following:

\begin{enumerate}
  \item Micro embeddings can support strong syntactic extension, exemplified by \texttt{fowl-geq}, where we are able to redefine the syntax for \texttt{J}, \texttt{Equal} and \texttt{refl} without modifying \texttt{fowl}.
  \item Micro embeddings can support additive rule-level semantic extension, since we are able to additively extend \texttt{fowl-base} with semantics for vectors and $\Sigma$ types in \texttt{fowl-vec} and \texttt{fowl-sigma}, implementing existing \texttt{fowl-base} judgements. However, when implementing \texttt{fowl-rett}, our implementation oversight means we are unable to replace \texttt{sort-term} without \textit{also} modifying judgements. Micro embeddings are therefore not shown to support strong rule-level semantic extension, but introducing interposition points for struct methods would resolve this issue.
  \item Micro embeddings can support strong judgement-level semantic extension, since we are able to add gradual types and judgements in \texttt{fowl-geq}, and also modify the original check judgements in \texttt{fowl-base}.
\end{enumerate}

\subsection{Performance}
\label{sec:performance}

To evaluate the performance of our micro embedding approach, we benchmark \texttt{fowl} against \texttt{cur} \cite{tsplus} (a dependently typed proof assistant written in the \texttt{turnstile+} ELF) \linebreak and against \texttt{smalltt} \cite{stt} (a small non-extensible high-\linebreak performance dependently typed language implementation) on a suite of dependently typed programs.

\texttt{cur} is a Racket hashlang, meaning that a \texttt{cur} program can be compiled to a Racket binary, performing all macro expansion into Racket core forms and type checking as specified in \texttt{cur}'s \texttt{turnstile+} implementation, resulting in an executable file. \texttt{fowl} behaves similarly, as it too is a Racket hashlang, compilation of which performs type checking and elaboration into Racket. Thus, \texttt{fowl} and \texttt{cur} benchmarking times are the complete compilation time including all parsing, expansion and compilation steps, including those of the Racket compiler.

In contrast, \texttt{smalltt} does not output compiled binaries, but does parse and type check programs. We use \texttt{smalltt} to represent the \textit{absolute best-case scenario} for performant ELFs like LMS and \texttt{syntax-spec}. \texttt{smalltt} benchmarks favorably against all mainstream dependently typed languages (being one to two orders of magnitude faster than \texttt{Agda} \cite{agda},	\texttt{Coq} \cite{coq}, \texttt{Lean} \cite{lean4}\footnote{\texttt{Lean} deserves note, as \texttt{Lean-4} has an extensible grammar which elaborates to a fixed core type theory referred to as the \texttt{Lean-4} \textit{kernel}. \texttt{Lean-4}'s extensions must all be transformed into a corresponding kernel representation, via an extensible elaborator from the surface grammar to the kernel language. \texttt{Lean-4} is thus akin to \texttt{syntax-spec} with a dependently typed back-end.}, and \texttt{Idris 2} \cite{Idris}, as seen in \Cref{fig:stttimes}) and uses both efficient term representation and optimizations to its normalization algorithm. Recall, \texttt{cur} has the additional work of re-walking macro expansions, and also has no algorithmic optimizations. \texttt{fowl} elaborates into an efficient IR representation before performing type checking (not needing to re-walk the AST), and performs elaboration in place, but otherwise does not make any efforts to avoid repeated work during normalization, and has no algorithmic optimizations. Any performance gains or losses relative to \texttt{cur} are thus predominantly due to \texttt{fowl}'s internal representation of language forms and hybrid embedding style. Since both \texttt{cur} and \texttt{fowl} share the same reader, runtime and compiler, their performance can be directly compared. This is not the case for \texttt{smalltt}, which we include only as a goalpost.

We use \texttt{hyperfine} \cite{hf} to estimate mean compilation time for \texttt{cur} and \texttt{fowl} across each benchmark. As configured, \texttt{hyperfine} reports the average duration of ten executions of \texttt{raco make}---which invokes the Racket compiler---on each benchmark program. Before these ten timed runs, three warm-up runs are performed and discarded to reduce the impact of cold processor caches. All benchmarks are executed in the same directory on the same 16-core Intel Skylake server machine with 16384 KB of L1 cache per core and 128 GB of total RAM. Each benchmark is executed sequentially, with a single core active at any one time. All benchmarking runs were completed within a 2-day period, with some \texttt{cur} runs not completing because the machine ran out of memory before type checking could complete. These runs are marked \textbf{DNF} in \Cref{fig:times}, with compilation times before failure all exceeding half an hour. For benchmarks of \texttt{smalltt}, we follow the procedure described in \citet{stt} and manually reload the benchmark in the \texttt{smalltt} repl thirteen times, discarding three warm-up runs. The mean and standard deviation in seconds for all runs is reported in \Cref{fig:times}.

\newcommand\perfentry[1]{\ifthenelse{\equal{#1}{0}}{\textbf{DNF}}{\tablenum{#1}{\,s}}}

\begin{table*}[h]
    \caption{Compilation times for \texttt{cur} and \texttt{fowl}, and load times for \texttt{smalltt} on a subset of the \texttt{smalltt} suite.}
  \label{fig:times}
\center
\begin{filecontents*}{data.csv}
name,cur,curstdev,fowl,fowlstdev,stt,sttstdev
asymp-small,29.12859398,0.050125507,0.783943211,0.005690109,0.002369707,0.000534173
asymptotics,0,0,0.966506545,0.007347008,0.025059095,0.003019183
conv-eval,1590.832745,2.70635184,4.896863992,0.022777241,0.001667979,0.000889616
stlc-small,362.103187,0.937433321,1.849867617,0.008437368,0.00171713,0.000153434
stlc-small5k,0,0,89.05711011,0.926582552,0.052548728,0.00154564
stlc-small10k,0,0,177.9284106,1.299219643,0.10308232,0.002808274
\end{filecontents*}
\csvreader[
  tabular = |r||c|c|c|c|c|c|,
  table head = \hline Benchmark & \texttt{cur} $\mu$ & \texttt{cur} $\sigma$ & \texttt{fowl} $\mu$ & \texttt{fowl} $\sigma$ & \texttt{smalltt} $\mu$ & \texttt{smalltt} $\sigma$\\\hline\hline,
  late after line = \\\hline,
  head to column names
  ]{data.csv}{}{
    \texttt{\name} &  \perfentry{\cur} & \perfentry{\curstdev} & \perfentry{\fowl} & \perfentry{\fowlstdev} & \perfentry{\stt} & \perfentry{\sttstdev}
  }
 \end{table*}

In \Cref{fig:timesplot} we plot mean compilation times for each language and benchmark. Note a clear trend across all benchmarks: \texttt{fowl} is about 1000x slower than \texttt{smalltt} across the board, and \texttt{cur} is 100x slower that \texttt{fowl}, when it could even compile the benchmark program. We discuss each benchmark in more detail below:

\begin{figure}[]
  \pgfplotsset{ compat=1.7 }
  \pgfplotstableread[col sep=comma]{data.csv}{\loadedtable}
  \center
  \begin{tikzpicture}
  \begin{semilogyaxis}[
      width=0.48*\textwidth,
      height=1.1*\axisdefaultheight,
      ylabel=Run time (s),
      ylabel shift = -0.4em,
      ymin=1e-3,
      ymax=1e4,
      ybar=0pt,
      xlabel=Benchmark,
      xlabel shift = -0.4em,
      xtick= {0,...,5},
      xticklabel style={rotate=30, anchor=north east},
      xticklabels from table={\loadedtable}{name},
      log origin=infty,
      bar width=1.2/6,
      enlarge x limits={abs=0.6},
      legend pos=north west,
      legend style={nodes={scale=0.8, transform shape}},
      legend cell align={left},
      nodes near coords={
          \pgfkeys{
              /pgf/fpu=true,
              /pgf/fpu/output format=fixed,
          }%
          \pgfmathparse{
            ifthenelse(
                \pgfplotspointmeta<1,
                \pgfplotspointmeta*1000,
                \pgfplotspointmeta
            )
        }
            \pgfmathtruncatemacro{\Y}{\pgfplotspointmeta}
        \ifnum\Y<1
            \pgfmathprintnumber{\pgfmathresult}\,ms
        \else
            \pgfmathprintnumber{\pgfmathresult}\,s
        \fi
      },
      nodes near coords style={
          font=\tiny,
          rotate=90,
          anchor=west,
      },
      point meta=rawy,
  ]
      \addplot [pattern=north east lines,
      pattern color=blue,
      error bars/.cd,
      y dir=both,
      y explicit]
      table [
          x expr=\coordindex,
          y=cur,
          y error=curstdev,
          col sep=comma,
      ] {\loadedtable};
      \addlegendentry{\texttt{cur} $\mu$};
      \addplot [pattern=north west lines,
      pattern color=red,
      error bars/.cd,
      y dir=both,
      y explicit]
      table [
        x expr=\coordindex,
        y=fowl,
        y error=fowlstdev,
        col sep=comma,
      ] {\loadedtable};
      \addlegendentry{\texttt{fowl} $\mu$};
      \addplot [pattern=crosshatch dots,
      pattern color=brown,
      error bars/.cd,
      y dir=both,
      y explicit]
      table [
        x expr=\coordindex,
        y=stt,
        y error=sttstdev,
        col sep=comma,
      ] {\loadedtable};
      \addlegendentry{\texttt{smalltt} $\mu$};
  \end{semilogyaxis}
  \end{tikzpicture}
  \caption{Compilation times for \texttt{cur} and \texttt{fowl}, and load times for \texttt{smalltt} on a subset of the \texttt{smalltt} benchmark suite. Omitted bars signify test execution failure.}
  \label{fig:timesplot}
  \vspace{1em}
\end{figure}
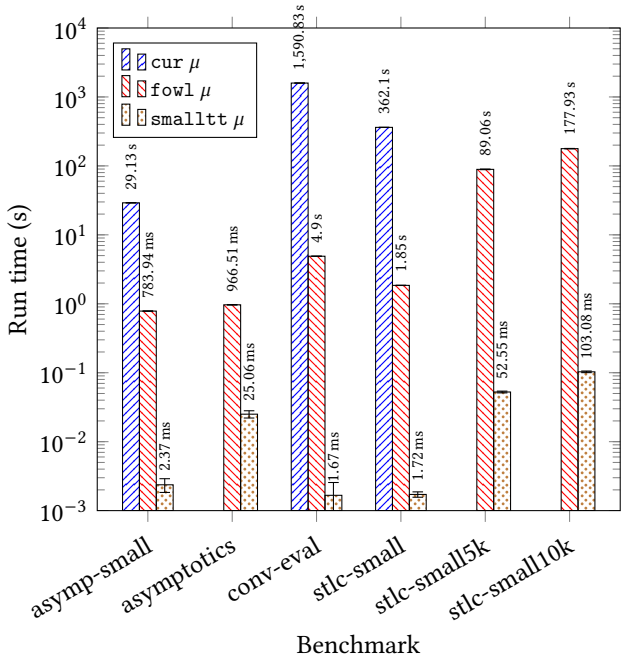

\paragraph{asymp-small} This benchmark defines an inductive vector type and constructs a 100-element vector, with each element being the base universe type, (\texttt{Type 0}). The \texttt{fowl} version of this benchmark uses the primitive vector defined in the \texttt{fowl-vec} extension; both the \texttt{cur} and \texttt{smalltt} versions use an inductively defined vector data type. \texttt{asymp-small} does not exist in the \texttt{smalltt} benchmark suite, and is a pared down version of the \texttt{asymptotics} benchmark. We include \texttt{asymp-small} since \texttt{cur} terminates without running out of memory on our benchmarking machine.

\paragraph{asymptotics} This benchmark is similar to \texttt{asymp-small}, but constructs a 1000-element vector. \texttt{smalltt} displays a nearly 10x increase in type checking time compared to \linebreak \texttt{asymp-small}, implying a roughly linear increase in algorithm run time with term depth. \texttt{fowl} took barely longer to compile the 1000 element vector than it did to compile \texttt{asymp-small}, and thus type checking did not constitute the majority of \texttt{asymp-small}'s compilation time. \texttt{cur} failed to compile the 1000 element vector with 128GB of available system memory. \texttt{asymptotics} is a subset of the \texttt{asymptotics} benchmark present in the \texttt{smalltt} repository, as it sufficed to demonstrate the divide in \texttt{cur}, \texttt{fowl} and \texttt{smalltt}'s performance.

\paragraph{conv-eval} This benchmark defines inductive naturals (using built-in types in \texttt{fowl}), addition and multiplication (using standard library functions in \texttt{cur}), then uses these functions to construct larger and larger natural numbers. The version of \texttt{conv-eval} in the \texttt{smalltt} repository constructs both larger numbers and has additional tree constructions, but \texttt{fowl} lacks the ability for us to define inductive data types and both \texttt{fowl} and \texttt{cur} failed to type check the larger numbers present in the original benchmark. \texttt{conv-eval} highlights the importance of \textit{glued evaluation}, a mechanism by which \texttt{smalltt} avoids normalization for syntactically equivalent terms. Neither \texttt{cur} nor \texttt{fowl} implement glued evaluation, and therefore \texttt{conv-eval} shows the largest performance divide between \texttt{fowl} and \texttt{smalltt}.

\paragraph{stlc-small} This benchmark first church-encodes a simply typed lambda calculus (\textit{STLC}) and type checks the function $\lambda x : (\bot \to \bot). \lambda y : \bot. x (x (x (x (x (x y)))))$. The benchmark file has a length on the order of 50-100 lines of code (\textit{loc}). All of \texttt{cur}, \texttt{fowl}, and \texttt{smalltt} perform as expected: \texttt{fowl} suffers compared to \texttt{smalltt}, since the complexity of church-encoded STLC terms is significant and \texttt{fowl} does not avoid the re-computation of normal forms. \texttt{cur} performs much worse, as the high term depth requires significant computational effort to reify into \texttt{cur} syntax after expansion. Both the \texttt{fowl} and \texttt{cur} benchmarks are written with explicit context types, as neither perform unification and type inference, unlike \texttt{smalltt}. \texttt{smalltt} thus performs additional work during type checking. \texttt{stlc-small} is unmodified from the \texttt{smalltt} benchmark suite.

\begin{table*}[h]
  \caption{Excerpt of reported type checking time of benchmarks in the \texttt{smalltt} benchmark suite in \citet{stt}, with scaled \texttt{fowl} and \texttt{cur} compilation times for comparison.}
\label{fig:stttimes}
\begin{tabular}{|r||c|c|c|c|c||c|c|}
  \hline
  Benchmark    & \texttt{smalltt} & \texttt{Agda} & \texttt{Coq} & \texttt{Lean} & \texttt{Idris 2} & \texttt{fowl} (x0.7) & \texttt{cur} (x0.7) \\ \hline \hline
   \texttt{stlc-small}    & \perfentry{0.003} & \perfentry{0.106} & \perfentry{0.128} & \perfentry{0.073} & \perfentry{0.542}  & \perfentry{1.295} & \perfentry{217.262} \\ \hline
   \texttt{stlc-small5k}  & \perfentry{0.037} & \perfentry{4.445} & \perfentry{0.762} & \perfentry{2.649} & \perfentry{6.397}  & \perfentry{62.353} & \perfentry{0} \\ \hline
   \texttt{stlc-small10k} & \perfentry{0.072} & \perfentry{22.80} & \perfentry{1.388} & \perfentry{5.244} & \perfentry{13.496} & \perfentry{124.550} & \perfentry{0} \\ \hline
\end{tabular}
\end{table*}

\paragraph{stlc-small5k} This benchmark executes 96 copies of \linebreak \texttt{stlc-small}, totaling on the order of 5000-10000 loc.
\texttt{fowl} and \texttt{smalltt} both take 50x longer on \texttt{stlc-small5k} than on \texttt{stlc-small}. \texttt{cur} runs out of memory and fails to compile the benchmark. \texttt{stlc-small5k} is unmodified from the \texttt{smalltt} benchmark suite.

\paragraph{stlc-small10k} This benchmark executes 192 copies of \texttt{stlc-small}.
\texttt{fowl} and \texttt{smalltt} both take 2x longer on \linebreak \texttt{stlc-small10k} than on \texttt{stlc-small5k}. This is ideal, as we see a linear increase in runtime over \texttt{stlc-small5k}. \texttt{cur} again runs out of memory and fails to compile the benchmark. \texttt{stlc-small10k} is unmodified from the \texttt{smalltt} suite.

\

We conclude that \texttt{fowl} performs about 10x slower than most modern dependently typed languages, far better than the 1000x slower performance of \texttt{cur}. We do so by comparing our benchmark runs for \texttt{smalltt} with the values reported in \citet{stt}, which can be seen in \Cref{fig:stttimes}. Our runs for \texttt{smalltt} are 30\% slower than those reported by \citet{stt}, so we include our measured compilation times for \texttt{cur} and \texttt{fowl} scaled by a factor of 0.7 for comparison. This performance is, however, given \textit{no effort} put towards optimizing \texttt{fowl}'s type checking algorithm, only by micro embedding \texttt{fowl} into an efficient struct based IR that allows us to perform IR elaboration in place. We expect implementing algorithmic optimizations will get the performance of \texttt{fowl} significantly closer to the state of the art, but doing so is left to future work. Crucially, \texttt{fowl} has roughly linear, not quadratic, increases in compilation time with increased term complexity, meaning that \texttt{fowl}'s term representation is not limiting object language performance, unlike \texttt{turnstile+}.

%% file: future.tex
\section{Future Work}
\label{sec:future}

Currently, our micro embedding approach is a series of design patterns.
This is, of course, antithetical to the Scheme ethos.
Instead, these design pattern should be wrapped up in some abstractions, ideally implemented using macros (what else?).
We conjecture we need three new abstractions, and have sketched an implementation of one.

\paragraph{micros} We can wrap up our mule pattern for implementing micros into its own \texttt{define-micro} form for defining micros.
This is easily done: \texttt{define-micro} should bind an identifiers in the transformer environment to a function from syntax to an arbitrary IR representation, such as structs.
It would automatically wrap the output in the mule pattern to smuggle the IR through the macro expander.
The micro abstraction should also provide a \texttt{expand-micros} form that locally expands until a fully expanded mule, like our \texttt{elab-to-structs} implementation in \Cref{fig:syndef} or like the custom expand function provided by \texttt{syntax-spec}.
The difficulty for us is in generating an extensible syntax class, called \texttt{expanded-term} in \Cref{fig:syndef}, but this is easily done with another pattern using a parameter that holds an open recursive syntax parser that is extended by each micro.
A prototype implementation is available as \texttt{syntax/micros} in the \materials.

\paragraph{extensible generics} Our pattern for extending judgements might be abstracted into extensible generics for structs.
We would define a form \texttt{extensible-generic} for defining new extensible generics, where the generic function indirects through an interposition point (a parameter).
Later, the generic could be updated with new functionality, even changing its arity.
Our experiments so far suggest this more general abstraction is more difficult to implement than our more limited pattern for extensible judgements.
Generics introduce a struct property that is attached to the struct, and this property enables accessing the generic function directly, not through our interposition.
The struct property itself cannot be easily interposed upon.
It might be that we instead need to introduce a more limited abstraction for extensible judgements.

\paragraph{extensible structs} The next abstraction we need is an extensible struct: structs that can be updated later to include new fields.
After updating, old users of the struct should implicitly create the new struct, and matching or projecting from the struct should work seamlessly, so this is not merely struct subtyping.
This would be necessary to support the RETT extension properly.
We would extend our design pattern for extensible judgements to structs: we would introduce an interposition point for each struct constructor, accessor, and match expander using a parameter.
Calls to the ``constructors'', \emph{e.g.}, would actually dereference the current constructor from the parameter.
A prototype is available in the \materials, although it does not support all necessary features, such as extending the matching expander.

%% file: acks.tex
\newcommand{\mynum}{RGPIN-2019-04207}

\begin{acks}
We gratefully acknowledge the feedback of the anonymous reviewers of this work, and the feedback of Ronald Garcia and Yuanhao Wei.

  We acknowledge the support of the \grantsponsor{GS501100000023}{Natural Sciences and
    Engineering Research Council of Canada (NSERC)}{http://dx.doi.org/10.13039/501100000023}, funding reference number
  \grantnum{GS501100000023}{\mynum} and CGS-M Award \#6563.
  We also acknowledge the support of the province of British Columbia, BC Graduate Scholarship \#6768.
  Cette recherche a été financée par le \grantsponsor{GS501100000023}{Conseil de
    recherches en sciences naturelles et en génie du Canada
    (CRSNG)}{http://dx.doi.org/10.13039/501100000023}, numéro de référence
  \grantnum{GS501100000023}{\mynum} et le prix CGS-M \#6563.
\end{acks}

%% file: data-availability.tex
\section*{Data Availability Statement}
The software artifacts, containing the implementation of \linebreak \texttt{fowl}, the test suite, and a prototype implementation of extensible micros and structs, have been made publically available~\cite{artifact}.